\documentstyle[aps,prb,epsfig,twocolumn,floats,psfig]{revtex}
\begin{document}
\draft
\twocolumn[\hsize\textwidth\columnwidth\hsize\csname @twocolumnfalse\endcsname

\title{Quantum critical point in a periodic Anderson model}
\author{Peter van Dongen$^{(1)\!}$, Kingshuk Majumdar$^{(2)\!}$ \cite{byline}, Carey Huscroft$^{(2)\!}$, and Fu-Chun Zhang$^{(2)\!}$}
\address{$^{(1)\!}$ Institut f\"{u}r Physik, Johannes Gutenberg-Universit\"{a}t, 55099 Mainz, Germany;\\ $^{(2)\!}$ Department of 
Physics, University of Cincinnati, OH 45221-0011}
\date{\today}
\maketitle

\begin{abstract}
{We investigate the symmetric Periodic Anderson Model (PAM) on a three-dimensional cubic lattice with nearest-neighbor hopping and 
hybridization matrix elements. Using Gutzwiller's variational method and the Hubbard-III approximation (which corresponds to the exact 
solution of an appropriate Falicov-Kimball model in infinite dimensions) we demonstrate the existence of a quantum critical point at 
zero temperature. Below a critical value $V_c$ of the hybridization (or above a critical interaction $U_c$) the system is an 
{\em insulator\/} in Gutzwiller's and a {\em semi-metal\/} in Hubbard's approach, whereas above $V_c$ (below $U_c$) it behaves like a 
metal in both approximations. These predictions are compared with the density of states of the $d$- and $f$-bands calculated from 
Quantum Monte Carlo and NRG calculations. Our conclusion is that the half-filled symmetric PAM contains a {\em metal-semimetal 
transition\/}, not a metal-insulator transition as has been suggested previously.}

\end{abstract}
\bigskip
]
\narrowtext

\section{Introduction}
\label{Introduction}

Heavy fermion systems, whose properties are determined by nearly localized, strongly correlated $f$-electrons hybridizing with the $d$-electrons
of the conduction band, have been of considerable interest in recent years.\cite{lee} The appropriate theoretical description of heavy fermion
systems is believed to be the Periodic Anderson Model (PAM). Among other properties of heavy fermions, this model explains the Kondo effect,
i.e., the quenching of the magnetic moments of the correlated electrons by the conduction electrons\cite{hewson}. Recently, using Quantum Monte
Carlo (QMC) techniques, Huscroft et al.~\cite{carey1,carey2} studied a Periodic Anderson model with the dispersion of the hybridization
proportional to that of the conduction electrons. Their study demonstrates that, as the temperature is lowered, the spin singlet correlation
function of the conduction electrons develops a sharp structure near a critical value of the hybridization. This indicates a very rapid
cross-over between a Kondo regime and a regime where the correlated electrons have unquenched moments. The nature of this crossover was also
addressed by Held and Bulla,\cite{HeldBulla} who showed that the Periodic Anderson model {\em under certain assumptions\/} contains a transition
equivalent to the Mott-Hubbard metal-insulator transition in the Hubbard model.\cite{Gebhard} Since it is clear already from Ref.\
\onlinecite{HeldBulla} that one of the assumptions (the strict separation of high- and low-energy scales) is at best only approximately
fulfilled, further analytical and numerical studies of this transition in the PAM are clearly called for.

In this paper we study the nature of this Mott-Hubbard-like transition analytically, using both Gutzwiller's variational method\cite{gutz,brink}
and the Hubbard-III approximation \cite{HubbardIII}. We demonstrate that at half-filling there is a quantum critical point as a function of the
hybridization strength which separates a Kondo regime from a phase, in which the $d$- and $f$-bands are {\em weakly coupled\/} (in the
Hubbard-III approximation) or even {\em completely decoupled\/} (in Gutzwiller's approach). The disappearance of the Kondo peak beyond the
quantum critical point is intimately connected to our choice of the model, with a hybridization strength that vanishes at the Fermi surface of
the conduction electrons. We also present new results from 3-dimensional QMC calculations in support of the existence of a quantum critical point.

The grand canonical Hamiltonian describing the two-band periodic Anderson model for hybridized $d$- and $f$-electrons is
\begin{eqnarray}
H &=& \sum_{{\bf k}\sigma} \epsilon_{\bf k}^{\phantom{\dagger}}d^\dagger_{{\bf k}\sigma}
      d_{{\bf k}\sigma}^{\phantom{\dagger}} + \sum_{{\bf k}\sigma}
  V_{\bf k}^{\phantom{\dagger}}(d^\dagger_{{\bf k}\sigma}f_{{\bf k}\sigma}^{\phantom{\dagger}}
   +f^\dagger_{{\bf k}\sigma}d_{{\bf k}\sigma}^{\phantom{\dagger}}) \nonumber \\
 &+& U\sum_{\bf i} \left(n_{{\bf i}f \uparrow}- \textstyle{1\over2}\right)
      \left(n_{{\bf i}f \downarrow}- \textstyle{1\over2}\right) \nonumber \\
 &+& \sum_{{\bf i}\sigma} \epsilon_f n_{{\bf i}f \sigma} - \mu \sum_{{\bf i}\sigma}
 \left(n_{{\bf i}f \sigma}+n_{{\bf i}d \sigma}\right)\, .
\label{H0}
\end{eqnarray}
Here $d^\dagger_{{\bf k}\sigma} (f^\dagger_{{\bf k}\sigma})$ and $d_{{\bf k}\sigma}^{\phantom{\dagger}} (f_{{\bf k}\sigma}^{\phantom{\dagger}})$
are the fermionic operators which create and destroy the $d$-($f$-)band electrons of momentum ${\bf k}$ and spin $\sigma$, and $n_{{\bf
i}f\sigma}^{\phantom{\dagger}}=f^\dagger_{{\bf i}\sigma} f_{{\bf i}\sigma}^{\phantom{\dagger}}$ is the number operator for the $f$-electrons of
spin $\sigma$ at site ${\bf i}$. Furthermore, $V_{\bf k}$ is the momentum dependent hybridization term between $f$- and $d$-electrons. Following
Refs.\ \onlinecite{carey1,carey2}, we consider the dispersion of the $d$-band and the mixing term $V_{\bf k}$  to be that of nearest neighbor
hopping on a three-dimensional simple cubic lattice (with unit lattice constant) whereas the $f$-band is taken to be dispersionless:
\begin{eqnarray}
\epsilon_{\bf k} &=& -2 t\left[ \cos k_x +\cos k_y + \cos k_z \right]\, , \label{ep}
 \\
V_{\bf k} &=& -2 V\left[ \cos k_x+\cos k_y+ \cos k_z\right]\, ,
\label{v_k} \\
\epsilon_f &=& 0\, .
\nonumber
\end{eqnarray}
Here $t$ and $V$ are the hopping matrix elements between the $d$-$d$- and $f$-$d$-bands respectively. In this paper we study the symmetric PAM in
which the chemical potential is $\mu=0$ and $\langle n_f \rangle = \langle n_d \rangle =1$.

As pointed out also in Refs.\ \onlinecite{carey1,carey2}, there are good reasons for replacing the usual {\em momentum independent\/}
hybridization, $V_{\bf k}=V$, by the momentum {\em dependent\/} hybridization (\ref{v_k}). It follows from elementary symmetry arguments that the
$f$- and $d$- orbitals are essentially orthogonal on the same site. The orthogonality of $f$- and $d$- orbitals on the same site implies that the
hybridization is predominantly built up from nearest and further neighbor contributions. Our choice (\ref{v_k}), which assumes only nearest
neighbor contributions to the hybridization, reflects this fact in the simplest possible manner. As shown below, this momentum dependence of the
hybridization has important consequences, in particular for physical properties beyond the quantum critical point.

This paper is organized as follows. First, in section \ref{Methods}, we introduce our two main methods for investigating the PAM, namely
Gutzwiller's variational method and the Hubbard-III approximation. Our variational results for the symmetric PAM are presented in section
\ref{Gutzwillerapproach}; our main finding is that the Gutzwiller approach predicts a Brinkman-Rice type metal-insulator transition. Next, in
section \ref{HubbardIIIapproach}, we study the PAM in the Hubbard-III approximation, which is equivalent to the exact solution of a
Falicov-Kimball model in infinite dimensions. The Hubbard-III solution displays rich behavior as a function of the on-site interaction $U$,
including a resonance at the Fermi level for weak coupling, a metal-semimetal transition at an intermediate-coupling quantum critical point, and
weakly coupled $d$- and $f$-bands at strong coupling. We then compare the results from the Gutzwiller and Hubbard-III approaches to
QMC-simulations of the PAM on a three-dimensional lattice and also to infinite-dimensional QMC-results and to calculations based on the Numerical
Renormalization Group (NRG, see section \ref{QMCresults}). Finally, in sections \ref{Discussion} and \ref{Summary}, respectively, we discuss and
summarize our results.

\section{Methods and models}
\label{Methods}

Traditionally, in particular in the context of the Hubbard model, \cite{Gebhard} there are two famous approaches for investigating
metal-insulator transitions, namely the variational approach pioneered by Gutzwiller \cite{gutz} and Brinkman and Rice,\cite{brink} and the Green
function decoupling scheme developed by Hubbard. \cite{HubbardIII} Both of these approaches are clearly approximate in nature. Gutzwiller's
method predicts the formation of an ever narrower quasiparticle peak, accompanied by a divergence of the effective mass, as the on-site
interaction $U$ approaches a critical value $U_{c}^{\rm Gutz}$ from below. For $U>U_{c}^{\rm Gutz}$, the Gutzwiller method leads to unphysical
results, such as the suppression of all hopping processes and all double occupancies. This method, therefore, is more realistic at weak than at
strong coupling. Hubbard's approximation, on the other hand, is generally considered to be more realistic at strong coupling. At weak coupling it
predicts the steady decrease of the number of charge carriers at the Fermi level; however, this mechanism is implemented in such a way that Fermi
liquid properties are violated. At strong coupling Hubbard's method predicts band splitting, i.e., the formation of a lower and an upper Hubbard
band. Combination of both methods (the Gutzwiller method at {\em weak\/} and Hubbard's approximation at {\em strong\/} coupling) has yielded
valuable information on the metal-insulator transition in the Hubbard model. Here we combine both methods in order to shed light on the nature of
the metal-insulator transition in the PAM.

To study the ground state properties of the half-filled PAM, Eq.\ (\ref{H0}), within the Gutzwiller approach, we follow the variational procedure
of Rice and Ueda.\cite{gutz,rice1,rice2} These authors considered the PAM with on-site (rather than nearest-neighbor) hybridization. The central
aspect of the Rice-Ueda approach is the suppression of doubly occupied $f$-states. The Gutzwiller-correlated wave function, $|\psi_G \rangle$, is
for the case of the PAM defined as
\begin{equation}
|\psi_G \rangle = P|\psi_0 \rangle\, ,
\end{equation}
where $|\psi_0 \rangle$ is the wave-function for the uncorrelated ($U=0$) ground state at half-filling, and $P$ is the Gutzwiller correlator,
defined as
\begin{equation}
P = g^{\hat D} = \prod_{\bf i} \left[1-\left(1-g\right){\hat D_{\bf i}} \right]\, . \label{pp}
\end{equation}
The operator ${\hat D}=\sum_{\bf i} n_{{\bf i}f \uparrow}n_{{\bf i}f \downarrow}$ in Eq.\ (\ref{pp}) is the double occupancy operator for the
$f$-electrons and $g$ is a variational parameter. For $g=0$, the operator $P$ projects all the states onto the subspace without doubly occupied
$f$-sites, whereas $g=1$ corresponds to the uncorrelated state. In general $g$ has to be determined by minimizing the total energy of the system
in the correlated state.

For the symmetric PAM, which is the case of interest in this paper, the Gutzwiller correlator $P$ is treated by renormalizing all hopping
processes by a Gutzwiller factor $q(\bar{d})$, where $\bar{d}=D/{\cal N}$ is the fraction of doubly occupied $f$-sites and ${\cal N}$ is the
total number of lattice sites. This approximation, which is alternatively referred to as the ``Gutzwiller approximation'' or as ``semi-classical
counting'', becomes {\em exact\/}\cite{GebhardPAM} (at least within the Gutzwiller variational approach, not for the PAM-Hamiltonian itself) in
the limit of high spatial dimensions ($d=\infty$).

The central element in Hubbard's Green function decoupling scheme is the so-called ``alloy analogy'', in which it is assumed that the down-spins
hop while the up-spins are immobile, and vice versa.\cite{resbroad} The Hubbard-III approximation, like the Gutzwiller approach, can be
understood as the exact solution of a simplified problem in high spatial dimensions. In the context of the PAM, the mobile nature of one spin
species (say the $f_{\downarrow}$-spins), interacting with an ``alloy'' of immobile electrons of opposite spin (here the $f_{\uparrow}$-spins),
can be described by the following Hamiltonian:
\begin{eqnarray*}
H &=& -t\sum_{\bf (ij)\sigma} d^\dagger_{{\bf i}\sigma}
      d_{{\bf j}\sigma}^{\phantom{\dagger}} - V\sum_{\bf (ij)}
  (d^\dagger_{{\bf i}\downarrow}f_{{\bf j}\downarrow}^{\phantom{\dagger}}
 +f^\dagger_{{\bf i}\downarrow}d_{{\bf j}\downarrow}^{\phantom{\dagger}}) \\
 &+& U\sum_{\bf i} \left(n_{{\bf i}f \uparrow}^{\phantom{\dagger}}-
  \textstyle{1\over2}\right)\left(n_{{\bf i}f \downarrow}^{\phantom{\dagger}}-
          \textstyle{1\over2}\right) \\
 &+& \sum_{{\bf i}\sigma}\epsilon_f^{\phantom{\dagger}}
      n_{{\bf i}f\sigma}^{\phantom{\dagger}}-\mu\sum_{{\bf i}\sigma}
 \left(n_{{\bf i}f\sigma}^{\phantom{\dagger}}
            +n_{{\bf i}d\sigma}^{\phantom{\dagger}}\right)\, .
\end{eqnarray*}
In this simplified model, the $f_{\uparrow}$-electrons form an alloy of immobile spins, since their hybridization with the $d$-band vanishes. As
a consequence, the $d_{\uparrow}$-electrons are completely decoupled from the rest of the system and can be integrated out. The creation
(annihilation) operators for the remaining $d_{\downarrow}$-electrons will simply be denoted by $d^\dagger_{\bf i}$ ($d_{{\bf
i}}^{\phantom{\dagger}}$) below. Since we consider the symmetric PAM ($\epsilon_f =0$) at half-filling ($\mu=0$), the last two terms in the
Hamiltonian drop out, and we are left with
\begin{eqnarray}
H &=& -t\sum_{\bf (ij)} d^\dagger_{\bf i}d_{\bf j}^{\phantom{\dagger}}
              -V\sum_{\bf (ij)}(d^\dagger_{\bf i}
                           f_{{\bf j}\downarrow}^{\phantom{\dagger}}
  +f^\dagger_{{\bf i}\downarrow}d_{\bf j}^{\phantom{\dagger}}) \nonumber \\
 &+& U\sum_{\bf i} \left(n_{{\bf i}f \uparrow}^{\phantom{\dagger}}
       - \textstyle{1\over2}\right)\left(n_{{\bf i}f \downarrow}^{\phantom{\dagger}}
          - \textstyle{1\over2}\right) \, .
\label{HSPAM}
\end{eqnarray}
This model is very similar to the spinless two-band Falicov-Kimball model, for which the Hubbard-III approximation (in the framework of the
Hubbard model) becomes exact in the limit $d\to\infty$. Analogously, the appropriate Green function decoupling (``alloy analogy'') for the PAM
becomes exact in the simplified model (\ref{HSPAM}) in high dimensions. Below, in section \ref{HubbardIIIapproach}, we solve the model
(\ref{HSPAM}) exactly in $d=\infty$. In order to obtain meaningful results in this limit, the hopping and the hybridization have to be scaled as
$t=\bar{t}/\sqrt{Z}$ and $V=\bar{V}/\sqrt{Z}$, respectively, where $Z$ denotes the number of nearest neighbors of each lattice site.

Part of the Hubbard-III approximation is the choice of the lattice. Hubbard \cite{HubbardIII} started from a semi-elliptical density of states
(DOS), which corresponds to the Bethe lattice in high dimensions. Below, we follow Ref.\ \onlinecite{HubbardIII} and solve (\ref{HSPAM}) on the
Bethe lattice. This choice has several advantages, e.g., that the bandwidth is finite in $d=\infty$, that the DOS near the band edges resembles
the DOS on a simple cubic lattice in $d=3$, and that the properties of the DOS can be studied {\em analytically\/}, since one obtains a
relatively simple closed equation for the local Green function.

Essentially the same Hamiltonian (\ref{HSPAM}), but now with on-site hybridization, was considered also by Consiglio and Gusm\~{a}o.
\cite{ConGus} These authors referred to the model as the Simplified Periodic Anderson Model (SPAM), a designation that we extend also 
to the case of more general (in particular nearest-neighbor) hybridization. The method of solution in Ref.\ \onlinecite{ConGus} was 
that of Brandt and Mielsch.\cite{BrMi} Here we use an alternative method\cite{PvDSHM} that is much better suited for the calculation of 
the DOS on the Bethe lattice. An early application of the Hubbard-III scheme to the periodic Anderson model, in particular a calculation
of the resistivity as a function of temperature, can be found in Ref.\ \onlinecite{Czycholl}.

\section{Gutzwiller's variational approach}
\label{Gutzwillerapproach}

In the Gutzwiller approximation, the kinetic energy terms of the PAM are mapped to an effective Hamiltonian with a renormalized hybridization.
This result was first obtained with the use of semiclassical counting arguments by Rice and Ueda\cite{rice1,rice2} (see also Ref.\
\onlinecite{vol}), and then put on a solid footing by Gebhard,\cite{GebhardPAM} who showed that this so-called ``Gutzwiller approximation''
becomes exact in the limit of infinite spatial dimensionality. We recall that Gutzwiller's variational scheme is equivalent to the slave-boson
mean-field theory of Kotliar and Ruckenstein\cite{KotRuck} at $T=0$. At half-filling, the effective Hamiltonian in the Gutzwiller approach
becomes\cite{rice1,rice2}
\begin{equation}
H_{\rm eff}=\sum_{{\bf k}\sigma} \epsilon_{\bf k}^{\phantom{\dagger}}
      d^\dagger_{{\bf k}\sigma}d_{{\bf k}\sigma}^{\phantom{\dagger}}
      + \sum_{{\bf k}\sigma} {\tilde {V}}_{\bf k}^{\phantom{\dagger}}
      (d^\dagger_{{\bf k}\sigma}f_{{\bf k}\sigma}^{\phantom{\dagger}}+
      f^\dagger_{{\bf k}\sigma}d_{{\bf k}\sigma}^{\phantom{\dagger}}) \; ,
\label{effH}
\end{equation}
with the renormalized hybridization
\[
{\tilde V}_{\bf k} = \sqrt{q(\bar{d})} \, V_{\bf k}\; .
\]
Here the renormalization factor $q(\bar{d})$ takes the form~\cite{rice1,rice2}
\[
 q = 8\bar{d}(1-2\bar{d})
\]
and $\bar{d}=D/{\cal N}$ is the fraction of doubly occupied $f$-sites. The ground state energy $E_g$ is now obtained from the expectation value of
$H_{\rm eff}$,
\[
E_g = \langle \psi_0|H_{\rm eff}|\psi_0 \rangle /{\cal N}+ U\bar{d}-U/4\, .
\]
The effective Hamiltonian is easily diagonalized. For convenience, we set $t=1$, to establish a unit of energy. One then finds two bands, with
eigenenergies $\vartheta_{\pm}\epsilon_{\bf k}$, where $\vartheta_{\pm}\equiv \frac{1}{2}(1\pm\sqrt{1+4qV^2})$. Note that there is no gap between
the two bands, and that both bands contribute to the ground state energy. This is in marked contrast to the results for on-site
hybridization.\cite{rice1,rice2} The result for the ground state energy is:
\begin{equation}
E_g = -|\epsilon_0|\sqrt{1+32\bar{d}(1-2\bar{d})V^2}+U\bar{d} - U/4 \, , \label{energy}
\end{equation}
where $\epsilon_0 = \sum_{\{ \epsilon_{\bf k}<0, \sigma\} } \epsilon_{{\bf k}\sigma}/{\cal N}$ is the ground state energy per lattice site of a
completely decoupled $d$-band ($U=V=0$). Minimizing the ground state energy with respect to $\bar{d}$ yields
\begin{equation}
\bar{d} = \frac{1}{4}\left\{ 1-\frac{U}{U_c}\sqrt{\frac{1+(2V)^2}{1+(2UV/U_c)^2}}\right\}\, . \label{d-eqn}
\end{equation}
Here we introduced the critical value $U_c$ of the interaction or, equivalently, a critical value $V_c$ of the hybridization, for which the
expectation value (\ref{d-eqn}) of the double occupancy vanishes:
\begin{equation}
U_c = 16V^{2}|\epsilon_0|\qquad ;\qquad V_c = \sqrt{\frac{U}{16|\epsilon_0|}} \, . \label{VcUc}
\end{equation}
Since $\bar{d}\geq 0$, Eq.\ (\ref{d-eqn}) only applies for $U\leq U_c$ (or $V\geq V_c$). For $U\geq U_c$, $\bar{d}=0$. Substituting Eq.\
(\ref{d-eqn}) into Eq.\ (\ref{energy}) gives the ground-state energy of the system:
\begin{equation}
E_g = -|\epsilon_0|\sqrt{\left[ 1+(2V)^2\right] \left[ 1+(2UV/U_c)^2\right]} \, . \label{energy2}
\end{equation}
We note that the variational ground-state energy is independent of the hybridization, $E_g = -\frac{U}{4} -|\epsilon_0|$, for $U>U_c$.
\begin{figure}[tb]
\protect \centerline{\epsfxsize=3.0in \epsfbox {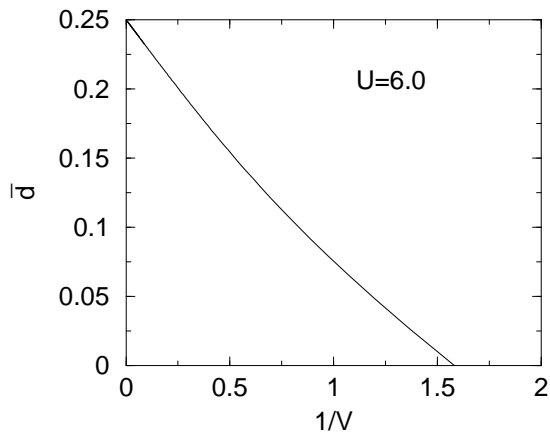}} \vskip .6cm \protect\caption{The fraction of doubly occupied sites is plotted as a
function of $1/V$, where $V$ is the hopping between the $f$- and the $d$-bands. For the critical $V_c$, given in Eq.\ (\ref{VcUc}), the number of
doubly occupied sites vanishes and all the sites become localized. The parameters are chosen to be $t = 1$ and $U = 6$, and the average energy of
the system is calculated by summing over all the filled bands: $|\epsilon_0| = 60/4^3$, where the lattice size is taken to be $4 \times 4 \times
4$. With the above values of the parameters the critical value of $V$ is approximately 0.63.} \protect\label{dt}
\end{figure}

In Fig.\ \ref{dt} we plot the fraction of doubly occupied $f$-sites $\bar{d}$ as a function of the inverse $d$-$f$-band hopping parameter $V$. The
fraction $\bar{d}$ decreases monotonically with $V$ and finally vanishes at a critical value $V_c$. This implies that, at this critical value of
$V$, every $f$-site is singly occupied, i.e., that the $f$-electrons are localized.

The critical hybridization strength $V_c$ (or, equivalently, the critical interaction $U_{c}$) defines a quantum critical point which separates
two distinct regimes in the model. For $V>V_c$ (or $U<U_c$), the $f$-electrons hybridize with the $d$-electrons and their moments are screened.
This is the Kondo regime. On the other hand, for $V<V_c$ (or $U>U_c$), the two bands are decoupled at the Fermi surface, and the $f$-electron
moments are unquenched. Thus the $f$-electrons show a ``metal-insulator transition'' from extended ($U<U_c$) to localized ($U>U_c$). In contrast,
the $d$-electrons at the Fermi level are always extended in both regimes.

Following Rice and Ueda,\cite{rice2} we introduce the {\em binding energy\/} in the Kondo regime as the singular part of the variational ground
state energy,
\[
E_{b}(U) \equiv -U/4 - |\epsilon_0| - E_{g}(U)\; ,
\]
so that $E_{b}(U)=0$ for all $U>U_c$. With this definition of the binding energy, it is easy to show that $E_{b}$ is positive for all 
$U<U_c$, and that for $U\uparrow U_c$ (i.e., if the transition is approached from below):
\[
E_{b}\sim \frac{2V^{2}|\epsilon_0|}{1+4V^2}\left( 1-\frac{U}{U_c}\right)^{2} \qquad (U\uparrow U_c) \; .
\]
Near the transition, where the two energy scales $E_{b}$ and $|\epsilon_0|$ are well-separated, it seems plausible that $E_{b}$ can be identified
with the Kondo temperature, $E_{b}\simeq k_{\rm B}T_{\rm K}$. It is of interest to compare our result for $T_{\rm K}$ to that of Rice and
Ueda,\cite{rice2} who found $T_{\rm K} = 2 |\epsilon_0| e^{-U/8V^2}$. While these results cannot be compared in detail (since Ref.\
\onlinecite{rice2} assumes a one-dimensional, {\em linear\/} dispersion for the $d$-electrons), it is nevertheless clear that the Kondo
temperature in our model is {\em strongly suppressed\/} relative to that of Ref.\ \onlinecite{rice2} for $U\lesssim U_c$.

The density of states of the PAM in the Gutzwiller approach is of interest, too. If we denote the DOS of a decoupled $d$-band ($U=V=0$) by $\nu
(E)$, then the DOS for the interacting $d$-electrons in the Gutzwiller approximation is simply given by
\[
\nu_{d}(E)=\frac{\nu (E/\vartheta_{+})+\nu (E/|\vartheta_{-}|)}{\vartheta_{+}+|\vartheta_{-}|}\; .
\]
Thus one finds that the DOS of the $d$-electrons near the transition ($U\lesssim U_{c}$) is strongly enhanced at the Fermi level compared to the
situation for $U>U_{c}$, where $\nu_{d}(E)=\nu (E)$. This result represents an interesting counter-example to the exhaustion scenario of
Nozi\`{e}res.\cite{noz} At low temperatures ($T<T_{\rm K}$) only the electrons within $T_{\rm K}$ of the Fermi surface can effectively
participate in screening the local moments. In a concentrated system such as ours, there are more moments to screen than conduction-band states
available for screening.  In the metallic regime of the conventional PAM, this should lead to a depletion of the density of screening states at
the Fermi surface.\cite{{niki1},{niki3}} However, within the Gutzwiller approximation this ``exhaustion physics'' is clearly absent, as we see an
enhancement of the $d$-band DOS near the Fermi surface.

Similarly the $f$-band DOS takes the form
\[
\nu_{f}(E)=\frac{(\vartheta_{+}/|\vartheta_{-}|)\nu (E/|\vartheta_{-}|)+(|\vartheta_{-}|/\vartheta_{+})\nu
(E/\vartheta_{+})}{\vartheta_{+}+|\vartheta_{-}|}\; .
\]
Thus, in the Gutzwiller approximation, the quasiparticle peak has width $|\vartheta_{-}|\propto q\propto \bar{d}\propto (1-U/U_{c})\propto
(|\epsilon_0|T_{\rm K})^{1/2}$, which is much larger than the Kondo scale $T_{\rm K}$. Hence, interestingly, the Kondo temperature $T_{\rm K}$
near the transition is {\em not\/} determined by the width of the peak in the $f$-DOS, but rather by the (much smaller) binding energy $E_{b}$.
The physical explanation for this is that screening becomes increasingly less efficient as one approaches the transition, due to the
renormalization of the hybridization rates, $\tilde{V}_{\bf k}=\sqrt{q}V_{\bf k}\to 0$ for $U\to U_c$.

\section{The Hubbard-III approach}
\label{HubbardIIIapproach}

The Hubbard-III approach (alloy analogy) is equivalent to the exact solution in $d=\infty$ of the Simplified Periodic Anderson Model (SPAM), Eq.\
(\ref{HSPAM}) with $t=\bar{t}/\sqrt{Z}$ and $V=\bar{V}/\sqrt{Z}$. For the calculation of the DOS on the Bethe lattice, it is most convenient to
first map the SPAM to an effective non-interacting Hamiltonian, following the lines of Ref.\ \onlinecite{PvDSHM}, and then use the renormalized
perturbation expansion \cite{Economou} to calculate the DOS. This method also allows one to conclude immediately that the DOS at half-filling is
{\em temperature independent\/}, as a consequence of the non-interacting nature of the effective Hamiltonian. Along the lines of Ref.\
\onlinecite{PvDSHM} we find that the Fourier transform of the local matrix Green function,
\[
G_{\bf ii}(\tau )= - \left( \begin{array}{c} \langle {\cal T} d_{\bf i}^{\phantom{\dagger}}(\tau )d^{\dagger}_{\bf i}(0)\rangle \;\;
   \langle {\cal T} d_{\bf i}^{\phantom{\dagger}}(\tau )f^{\dagger}_{{\bf i}\downarrow}(0)\rangle \\
\langle {\cal T} f_{{\bf i}\downarrow}^{\phantom{\dagger}}(\tau )d^{\dagger}_{\bf i}(0)\rangle \;\;
   \langle {\cal T} f_{{\bf i}\downarrow}^{\phantom{\dagger}}(\tau )f^{\dagger}_{{\bf i}\downarrow}(0)\rangle \end{array}\right)\; ,
\]
satisfies a cubic matrix equation,
\begin{equation}
G(z)=\textstyle{1\over2}\left\{ \left[ I_{-}-\Theta G\Theta\right]^{-1}+ \left[ I_{+}-\Theta G\Theta\right]^{-1}\right\}\; , \label{cubiceqG}
\end{equation}
with
\[
I_{\pm}(z)=\left( \begin{array}{cc} z & 0 \\ 0 & z\pm \frac{1}{2}U \end{array} \right) \qquad ;\qquad
           \Theta =\left( \begin{array}{cc} \bar{t} & \bar{V} \\ \bar{V} & 0 \end{array} \right)\; .
\]
In spite of the formally simple structure of Eq.\ \ref{cubiceqG}, the detailed analysis of the DOS of the $d$- and $f$-electrons is rather
involved and will be published elsewhere.\cite{PvDdetails} Here we focus on the physical content of the Hubbard-III results and compare them to
results from the Gutzwiller approach and various numerical techniques. For convenience we put $\bar{t}=1$ to fix the unit of energy; note that
this convention differs from that of the previous section.

\subsection{Density of states at strong interaction}
\label{DOSatlargeU}

First we consider the results from the Hubbard-III approximation in the limit of large interaction, $U\to\infty$. In this limit almost all the
spectral weight of the $d$-electrons is contained in a semi-elliptic band near $E=0$:
\begin{equation}
\nu_{d}(E)\sim \frac{1}{2\pi}\sqrt{4-E^2}\; .
\label{dDOSUtoinfty}
\end{equation}
Similarly, nearly all spectral weight of the
$f_{\downarrow}$-electrons is contained in two high and narrow peaks near $E=\pm\frac{1}{2}U$, whose width is of order $\bar{V}^{2}/U$ while
their height is proportional to $U/\bar{V}^{2}$. With a redefinition of the energy variable as $\lambda =U(|E|-\frac{1}{2}U)/\bar{V}^{2}$, one
finds
\[
\nu_{f\downarrow}(E)=\frac{U}{4\pi \bar{V}^{2}\lambda}\sqrt{6\lambda -1-\lambda^{2}}\; ,
\]
provided that the argument of the square root is positive ($3-\sqrt{8}<\lambda <3+\sqrt{8}$); otherwise $\nu_{f\downarrow}(E)$ vanishes. The
presence of an upper and a lower Hubbard band in the $f_{\downarrow}$-spectrum at first sight suggests the occurrence of a metal-insulator
transition at some finite value of $U$. However, closer inspection shows that there is some small {\em additional\/} spectral weight for the
$d$-electrons near $E=\pm\frac{1}{2}U$,
\[
\nu_{d}(E)\sim \frac{1}{2\pi U}\sqrt{6\lambda -1-\lambda^{2}}\; ,
\]
and small {\em additional\/} spectral weight for the $f_{\downarrow}$-electrons near $E=0$:
\[
\nu_{f\downarrow}(E)\sim \frac{2\bar{V}^{2}}{\pi U^{2}}E^{2}\sqrt{4-E^2}\; .
\]
This is a first clear indication that the $f_{\downarrow}$-electrons in the SPAM do {\em not\/} undergo a metal-insulator transition. Instead one
finds a {\em metal-semimetal transition\/}: at arbitrarily large $U$, the $d$- and $f_{\downarrow}$-bands remain weakly hybridized.

\subsection{Density of states at the Fermi level}
\label{DOSatFermilevel}

Exactly at the Fermi level ($E=0$) an explicit non-perturbative solution can be obtained for all interaction strengths $U$. A detailed analysis
of the cubic matrix equation (\ref{cubiceqG}) shows that there is only one physically acceptable solution,
\[
G(0)=\left( \begin{array}{cc} g_{1}(0) & g_{2}(0) \\ g_{2}(0) & g_{3}(0) \end{array} \right)\; ,
\]
where the matrix elements of $G(0)$ are given by
\begin{eqnarray*}
g_{1}(0) &=& -\sqrt{2}iR_{+}\\
g_{2}(0) &=& \frac{i}{\sqrt{2}\bar{V}}\left( R_{+}-R_{-}\right) \\
g_{3}(0) &=& -\frac{i}{\sqrt{2}\bar{V}^2}\left( R_{+}-R_{-}\right) \; ,
\end{eqnarray*}
and
\[
R_{\pm}\equiv\sqrt{1-\frac{U^{2}}{8\bar{V}^{4}} \pm \sqrt{1-\frac{U^{2}}{4\bar{V}^{4}}}} \; .
\]
These results hold only for $U\leq U_{c}^{\rm AA}=2\bar{V}^{2}$, where the superscript ``AA'' stands for ``alloy analogy''. For all $U>U_{c}^{\rm
AA}$ the solution is simply given by $g_{1}(0)=-i$ and $g_{2}(0)=g_{3}(0)=0$. Note that the Hubbard-III and Gutzwiller approaches predict the same
dependence of $U_c$ on the hybridization: both $U_{c}^{\rm AA}$ and $U_{c}^{\rm Gutz}$ are simply proportional to $V^2$.

Physically these results mean, that at weak coupling the DOS at the Fermi level decreases fairly slowly as the interaction $U$ is switched on,
both for the $d$- and the $f_{\downarrow}$-electrons:
\[
\left.
          \begin{array}{l} \nu_{d}(0)\sim \frac{2}{\pi}\left( 1-\frac{U^{2}}{16\bar{V}^{4}}\right)   \\
                 \nu_{f\downarrow}(0)\sim \frac{1}{\pi \bar{V}^{2}}\left( 1-\frac{U^{2}}{8\bar{V}^{4}}\right)\end{array}
\right\} \;\; (U\downarrow 0)\; ,
\]
while the critical values are approached quite rapidly:
\[
\left.
          \begin{array}{l} \nu_{d}(0)\sim \frac{1}{\pi}\left[ 1 + \sqrt{U_c^{\rm AA}-U}/\bar{V}\right]   \\
                 \nu_{f\downarrow}(0)\sim \frac{1}{\pi}\sqrt{U_c^{\rm AA}-U}/\bar{V}^{3}\end{array}
\right\} \;\; (U\uparrow U_{c}^{\rm AA})\; .
\]
For all $U\geq U_{c}^{\rm AA}$ the $f_{\downarrow}$-DOS at the Fermi level vanishes exactly, while the $d$-DOS is pinned at the value
$\nu_{d}(0)=1/\pi$. This clearly demonstrates that $U_c^{\rm AA}$ marks a quantum critical point, although the {\em nature\/} of the transition
cannot be deduced from an investigation of the DOS only at the Fermi level.

We add that, at weak coupling, the DOS for both the $d$- and the $f$-electrons displays an interesting resonance at the Fermi level. This
resonance is very narrow, of ${\cal O}(U^2)$, and its height remains of ${\cal O}(1)$ for $U\to 0$. The amplitude of the resonance for the
$d$-electrons is positive, so that the $d$-band DOS at $U=0^{+}$ is {\em larger\/} than for $U=0$. In contrast, the amplitude of the resonance in
the $f$-band is negative, i.e., the $f$-band DOS at $U=0^{+}$ is {\em smaller\/} than for $U=0$.

\subsection{Density of states at and beyond the critical point}
\label{DOSat&beyondUc}

We now consider the shape of the DOS as a function of energy, first for $U=U_c^{\rm AA}$ and then for $U>U_c^{\rm AA}$. We focus on the energy
interval near the Fermi level, since this interval determines the nature of the quantum critical point and virtually all physical properties of
interest.

First we present the result for the DOS exactly at the transition, i.e., for $U=U_c^{\rm AA}$. In this case one finds that the DOS has a sharp
(cubic-root) singularity as a function of energy for $|E|\to 0$:
\[
\left.
          \begin{array}{l} \nu_{d}(E)\sim \frac{1}{\pi}+\frac{\sqrt{3}}{2}\left( 1+\frac{1}{\bar{V}^{2}}\right)^{1/3}|E|^{1/3}   \\[2ex]
                 \nu_{f\downarrow}(E)\sim \frac{\sqrt{3}}{2\pi \bar{V}^{2}}\left( 1+\frac{1}{\bar{V}^{2}}\right)^{1/3}|E|^{1/3}\end{array}
\right\} \;\; (|E|\to 0)\; ,
\]
Cubic-root critical behavior of the DOS near the Fermi level is well-known from the Hubbard-III solution for the Hubbard model. Here we find that
similar critical behavior occurs also for the SPAM with nearest-neighbor hybridization.

For energies near the Fermi level, one generally finds that the $d$-DOS for $U>U_c^{\rm AA}$ is {\em metallic\/} while the $f_{\downarrow}$-DOS
is {\em semimetallic\/}. In particular, one finds that the $f_{\downarrow}$-DOS contains spectral weight, arbitrarily close to the Fermi level,
\[
\nu_{f\downarrow}(E)\sim \frac{4\bar{V}^{2}}{\pi U^{2}}E^{2}  \qquad (|E|\to 0, U>U_c^{\rm AA})\; .
\]
Note that the coefficient in front of the semimetallic $E^2$-behavior is valid for {\em all\/} $U>U_c^{\rm AA}$ and, hence, simply {\em
identical\/} to that calculated in section \ref{DOSatlargeU} for large $U$.

\section{Comparison to QMC and NRG results}
\label{QMCresults}

To obtain a complete picture of our model, we compare our analytical results to single-particle spectra obtained from three different numerical
techniques: ($i$) QMC simulations of the PAM on a three-dimensional lattice (where we present new results and compare to results from
recent\cite{carey1,carey2} calculations), ($ii$) QMC simulations of the PAM on an infinite-dimensional lattice\cite{carey2}, and ($iii$) recent
results from NRG calculations in infinite dimensions.\cite{HeldBulla}

\begin{figure}[tb]
\protect \centerline{\epsfxsize=3.5in \epsfbox {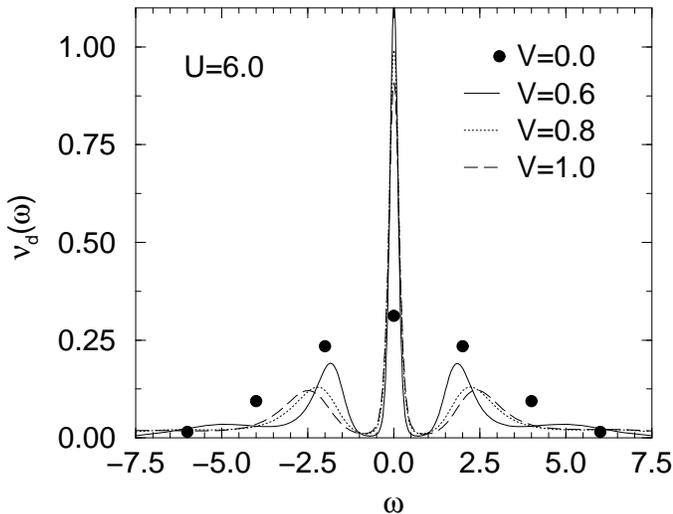}} \vskip .4cm \protect\caption{The $d$-band density of states, 
$\nu_{d}(\omega)$, for the $3d$ PAM via finite-$d$ QMC for various hybridizations, $V>V_c$. From Refs.\ \protect\onlinecite{carey1}
and \protect\onlinecite{carey2} we estimate that $V_{c}\simeq 0.5$. The DOS at the Fermi energy is resonantly enhanced over the 
non-interacting value, shown as dots in the figure. For $V<V_c$, the Green function is similar to the non-interacting Green function, 
which indicates that the $d$- and the $f$-bands are approximately decoupled in this regime (not shown).} \protect \label{dd}
\end{figure}

We start with the comparison to the three-dimensional QMC-results. In Fig.\ \ref{dd} we plot new results for the $d$-band DOS obtained using the
maximum entropy method to analytically continue three-dimensional QMC data calculated at $T=0.2$.  For small $V$ the presence of upper and lower
Hubbard bands well separated by a gap $U$ can be seen in the $f$-band DOS. With increasing $V$, the weight in the central region is enhanced at
the expense of the Hubbard side-bands.

For $V$ larger than the critical hybridization $V_c$, the DOS for the $d$-band is greatly enhanced compared to the non-interacting $d$-band DOS
(shown in dots).  This is consistent with the previously mentioned enhancement of the $d$-band DOS in the Gutzwiller approximation in this regime.
On the other hand, in the regime where $V<V_c$, the Green function is not significantly different from the non-interacting Green function. This
indicates that for $V<V_c$, the $d$-band is weakly coupled to the $f$-band. This result is compatible with both the Gutzwiller result (that the
$d$-band is strictly $V$-independent for $V<V_c$, i.e., $\nu_{d}(E)=\nu (E)$) and the Hubbard-III result, Eq.\ \ref{dDOSUtoinfty}, which holds
asymptotically both for $U\to\infty$ and $V\to 0$.

The $f$-band DOS was studied in Refs.\ \onlinecite{carey1,carey2} with the use of finite- and infinite-dimensional QMC techniques. The
infinite-dimensional (DMFT) calculations\cite{carey2} were also done for a strictly paramagnetic state, without antiferromagnetic fluctuations.
For small hybridization, $V<V_c$, one finds that the $f$-band DOS consists primarily of an upper and a lower Hubbard band around $E=\pm
{1\over2}U$, respectively. Only in the DMFT calculations\cite{carey2} is there a sign of small spectral weigth near the Fermi level. For
increasing $V$, one finds that the upper and lower Hubbard bands merge. Before the bands merge ($V\lesssim V_c$), additional resonant peaks
develop, which are characteristic for Kondo singlet formation. After the bands merge ($V>V_c$), the $f$-band DOS in the finite-dimensional
simulation shows some depletion near the Fermi level. Since this depletion is absent in the DMFT calculations, this feature is most likely due to
short-range antiferromagnetic fluctuations. Comparing these numerical results to those from the Hubbard-III and Gutzwiller approaches, we observe
that the merging of the two bands is described by the Hubbard-III approach but not by the Gutzwiller approximation. The Hubbard-III approach also
explains the small additional spectral weigth near the Fermi level in the DMFT calculation. The Gutzwiller approximation cannot explain either
the upper and lower Hubbard bands or small spectral weigth near the Fermi level. Obviously, neither approach can explain the depletion due to
short-range antiferromagnetic fluctuations in the finite-dimensional calculations.

We add that, very recently, Held and Bulla\cite{HeldBulla} performed calculations for the paramagnetic ground state of the PAM using the Numerical
Renormalization Group (NRG). These calculations clearly reveal significant additional spectral weight near the Fermi level for $U>U_c$. The
numerical results in Ref.\ \onlinecite{HeldBulla} are the clearest evidence to date that the quantum critical point in the PAM corresponds to a
metal-semimetal and not to a metal-insulator transition. Interestingly, the authors of Ref.\ \onlinecite{HeldBulla} show analytically that the
PAM {\em under certain assumptions\/} contains a transition equivalent to the Mott-Hubbard {\em metal-insulator\/} transition in the Hubbard
model. However, they also point out that one of these assumptions (the strict separation of high- and low-energy scales) is at best only
approximately fulfilled for $U\gtrsim U_c$, especially for larger values of the hybridization rate ($V\gtrsim 1$). The Hubbard-III results of
this paper strongly suggest that the high- and low-energy scales in the PAM are in fact never strictly separated, so that the transition is from
a metal to a semimetal, not an insulator. Physically it seems obvious that for large $U$, as a result of the finite hybridization rate, the
$f$-band must have some spectral weight near the Fermi level, where the weight of the $d$-band is concentrated. Similarly, the $d$-band must have
some spectral weight near the upper and lower Hubbard bands of the $f$-electrons, i.e., near $E=\pm {1\over2}U$.

\section{Discussion}
\label{Discussion}

The two main questions to be discussed are: What can be learned from the Gutzwiller and Hubbard-III approximations for the PAM? And what features
are missing in these two analytical approximation schemes?

As pointed out before, in section \ref{Methods}, one expects Gutzwiller's variational method to be more realistic at weak than at strong
coupling. This expectation is based on experience with the Hubbard model and on the fact that the Gutzwiller wave function becomes exact
in the weak coupling limit. For similar reasons, Hubbard's approximation is considered to be more realistic at strong coupling. For
example, the Hubbard-III approximation violates weak-coupling (Fermi-liquid) properties and is, therefore, better suited for
$U>U_c$, where the DOS of the $f$-electrons displays essentially no low-energy spectral weight.\cite{logic} Combination of these 
two approaches leads to the following physical picture.

We distinguish high and low spatial dimensions. In high spatial dimensions, short-range antiferromagnetic fluctuations (in particular
due to spin-flip processes on nearest-neighbor sites) are small, which is a prerequisite for the validity of the Gutzwiller and 
Hubbard-III approaches. The Gutzwiller approach then shows that the $f$-band DOS develops a quasiparticle peak near the Fermi level if 
the interaction $U$ is turned on. The quasiparticle peak becomes ever narrower, until it vanishes at the critical interaction $U_c$, 
where the metal turns into a semi-metal. The Gutzwiller approach is unable to describe the side bands in the DOS that develop for 
$U\lesssim U_c$. These high-energy features are better captured by the Hubbard-III approximation which, in turn, is unable to
describe the Fermi-liquid peak. For $U\gtrsim U_c$ one expects the DOS to remain semi-metallic, as is seen in the Hubbard-III approach. 
The Hubbard-III results further suggest that, at some finite interaction $U>U_c$, the DOS splits into three parts: a lower Hubbard band 
near $E=-{1\over2}U$, a semimetallic low-energy part of the spectrum, and an upper Hubbard band near $E=+{1\over2}U$. This scenario is 
consistent with the results from the high-dimensional QMC simulations and agrees very well with those of the NRG calculations discussed 
in section \ref{QMCresults}.

In contrast, there will be significant short-range antiferromagnetic fluctuations in the paramagnetic phase in low spatial dimensions (certainly
for $d=1,2$ and to some extent also for $d=3$). In these dimensions one, therefore, expects depletion of the $f$-band DOS near the Fermi level.
This depletion is explicitly seen in three-dimensional QMC-calculations.\cite{carey1,carey2} These depletion effects cannot be described in terms
of the paramagnetic Gutzwiller wave function for the PAM, since it does not contain the necessary antiferromagnetic fluctuations. The effects of
finite-dimensionality on the quality of the Hubbard-III results for $U>U_c$ is probably not large. One expects the semimetallic part of the DOS
to be somewhat depleted in $d=3$ as compared to $d=\infty$. An interesting question that cannot be answered at this stage is, whether the
semimetallic pseudogap changes its functional form due to short-range antiferromagnetic fluctuations and, in particular, whether these
fluctuations might turn the pseudogap into a hard gap. Given these uncertainties concerning the influence of antiferromagnetic fluctuations in
finite dimensions, it seems fair to say that our results are at least consistent with the three-dimensional QMC data published here and in Refs.\
\onlinecite{carey1} and \onlinecite{carey2}. We do not believe that our results can reasonably be applied to one- and two-dimensional systems,
since physics in these low dimensions differs too much from that in high dimensions, where our approximations are valid.

In the Gutzwiller and Hubbard-III methods studied in this paper, only states without broken symmetry were considered. The true ground 
state of the half-filled symmetric PAM, of course, may well be antiferromagnetically ordered.\cite{carey1,carey2,mark,coleman} In this 
sense our results are not so much relevant for the ground state of the PAM, but rather for the paramagnetic phase at slightly elevated 
temperatures ($T\gtrsim T_{\rm N\acute{e}el}$). Note that the low-temperature antiferromagnetic phase can be partially or entirely
suppressed due to frustration.\cite{frustration} With this proviso, we believe that the main result of our paper, the absence of Kondo 
screening for $V<V_c$ (or $U>U_c$), is robust. In particular, we expect that metal-semimetal transition, predicted by the Hubbard-III 
approach, actually occurs in the PAM with nearest-neighbor hybridization.

A comment is in order concerning the critical values of the interaction in the Gutzwiller and Hubbard-III (``alloy analogy'') approaches,
$U_{c}^{\rm Gutz}$ and $U_{c}^{\rm AA}$. In sections \ref{Gutzwillerapproach} and \ref{HubbardIIIapproach} we found that $U_{c}^{\rm
Gutz}=16V^{2}|\epsilon_0|/t^2$ and $U_{c}^{\rm AA}=2\bar{V}^{2}/\bar{t}$, where we reinstated factors of $\bar{t}$ and $t=\bar{t}/\sqrt{Z}$. In
order to be able to compare these two results, we calculate the critical value $U_{c}^{\rm Gutz}$ for the Bethe lattice with coordination number
$Z\to\infty$. One readily finds that in this case $|\epsilon_0|=8\bar{t}/3\pi$, so that $U_{c}^{\rm Gutz}=128\bar{V}^{2}/(3\pi\bar{t})$. Hence
there is a significant discrepancy between the two $U_c$-values: $U_{c}^{\rm Gutz}/U_{c}^{\rm AA}=64/3\pi\simeq 6.79$. Partly this is due to the
neglect of ``resonance broadening corrections'', which are part of the full Hubbard-III approximation but are not taken into account in the
alloy-analogy approach.\cite{resbroad} For the Hubbard model, the resonance broadening corrections are known\cite{Gebhard} to enhance $U_{c}^{\rm
AA}$ by a factor of $\sqrt{3}$, so that one then finds a ratio $U_{c}^{\rm Gutz}/(\sqrt{3}U_{c}^{\rm AA})=32/3\sqrt{3}\pi\simeq 1.96$. Hence the
Gutzwiller prediction for $U_c$ in the Hubbard model is significantly larger than the full Hubbard-III result. In the PAM the discrepancy is even
larger. Even if resonance broadening corrections are taken into account through an {\em ad hoc\/} factor of $\sqrt{3}$, $U_{c}^{\rm Gutz}$ is
still larger than $\sqrt{3}U_{c}^{\rm AA}$ by a factor of nearly 4. In order to determine which of the three predictions $U_{c}^{\rm Gutz}$,
$U_{c}^{\rm AA}$ and $\sqrt{3}U_{c}^{\rm AA}$ is best, we compare with results from recent NRG-calculations for the ground state of the PAM in
high dimensions.\cite{HeldBulla} E.g., for $\bar{t}=\bar{V}=1$ one finds that $U_c^{\rm PAM}\simeq 4$. In conclusion, compared to the NRG-value,
$U_{c}^{\rm AA}$ is too small by a factor of 2 due to the neglect of resonance broadening corrections, the ``corrected'' value
$\sqrt{3}U_{c}^{\rm AA}$ is quite close, and the Gutzwiller prediction $U_{c}^{\rm Gutz}$ is too large by a factor of $\simeq 3.395$. For the
PAM, therefore, the full Hubbard-III approximation leads to a much more accurate prediction for the critical interaction than the Gutzwiller
approach.

We now comment on the influence of small deviations from the strict nearest-neighbor hybridization, $V_{\bf k}^{\rm es}=-2 V\sum_{\ell =1}^{d}
\cos (k_\ell )$, which we considered throughout in this paper. As an example, we consider combinations of this ``extended $s$-wave'' form and a
local (``$s$-wave'') hybridization, $V_{\bf k}^{\rm s}=-V_{\rm s}$, or a ``$p$-wave'' form, $V_{\bf k}^{\rm p}=-2 V_{\rm p}\sum_{\ell =1}^{d}
\sin (k_\ell )$. In the Gutzwiller approximation, one readily finds that there is now a gap at the Fermi surface and that the renormalization
factor $q$ is now {\em finite\/} for all $U>0$. For instance, one finds for a mixture of an extended $s$-wave and a $s$-wave:
\begin{equation}
\label{sphybr} q\propto \frac{t^2}{V_s^2}\exp\left[ -\frac{U-U_{c}(V)}{16\nu (0)V_{s}^{2}}\right]\qquad (U\gtrsim U_c)\; .
\end{equation}
For a mixture of an extended $s$-wave and a $p$-wave, the factor $V_{s}^{2}$ in the exponent is replaced by the average of $\left( V_{\bf k}^{\rm
p}\right)^{2}$ over the non-interacting Fermi surface of the $d$-electrons. Strictly speaking, Eq.\ (\ref{sphybr}) implies that the quantum
critical point, which occurs for $V_{\rm s}=V_{\rm p}=0$, is unstable with respect to small $s$- or $p$-wave perturbations. However, Eq.\
(\ref{sphybr}) also shows that the QCP at $U_{c}(V)=16V^{2}|\epsilon_0|/t^2$ is replaced by a sharp crossover to a state with very heavy
quasiparticles. Experimentally this could hardly be distinguished from the metal-insulator (or metal-semimetal) transition found for $V_{\rm
s}=V_{\rm p}=0$.

Finally we comment on the difference between the local and nearest-neighbor hybridization rates from a renormalization group point of view. The
Kondo effect is due to the antiferromagnetic coupling of $d$-electrons near the Fermi surface to the $f$-electron moments. As the system
renormalizes, the coupling scales to the strong coupling ($J=\infty$) fixed point. Adjacent to this fixed point, only energy scales and momenta
near the Fermi surface are important.  For the standard PAM the hybridization is constant, and one finds the usual Kondo effect. In contrast, the
most salient feature of the model we study is that the $f$-$d$ hybridization {\em vanishes\/} at the Fermi surface of the half filled
non-interacting model, as can be seen from Eqs.\ (\ref{ep}) and (\ref{v_k}). The corresponding exchange coupling also vanishes on the Fermi
surface, and therefore cannot be rescaled to the strong coupling limit. Thus the absence of the Kondo effect in our model at small values of
$V/t$ is fully consistent with the usual renormalization group arguments.

\section{Summary}
\label{Summary}

We have studied the half-filled symmetric Periodic Anderson model with an $f$-$d$ hybridization proportional to the $d$-band dispersion, using
Gutzwiller's variational wave function and Hubbard's alloy-analogy approach. Both methods demonstrate the occurrence of a quantum phase
transition for a critical value $V_c$ of the $f$-$d$ hybridization rate or, equivalently, for a critical interaction strength $U_c$. For $V>V_c$
(or $U<U_c$) the system is a metal. For $V\leq V_c$ (or $U\geq U_c$) the system behaves as an insulator in Gutzwiller's approximation and as a
semi-metal in the Hubbard-III approach. Based on the results of these two approximate methods, we predict the occurrence of a similar quantum
phase transition in the exact solution of the PAM with nearest-neighbor hybridization. A number of the properties of this transition, such as a
strong suppression of the Kondo temperature on the metallic side, were also predicted. Most importantly, we conclude from our results that the
{\em nature\/} of the transition in the exact solution of the PAM will be that of a {\em metal-semimetal\/}, not a metal-insulator\cite{HeldBulla}
transition. These analytical results are compared with the DOS obtained from QMC-simulations and NRG-results for the $d$- and $f$-bands. We find
good agreement between our scenario and QMC- or NRG-results in high spatial dimensions ($d=\infty$), while our results are at least consistent
with the QMC-calculations published here and in Refs.\ \onlinecite{carey1} and \onlinecite{carey2} for the three-dimensional PAM.

\acknowledgements We acknowledge important and very helpful discussions with Prof.\ Mark Jarrell (Univ.\ of Cincinnati). This work was supported
by PRF (KM, FZ) and the NSF grant DMR-9704021 (CH, KM). We thank the U.S. Dept. of Energy ASCI program for supercomputer time for the QMC
simulations.


\begin{thebibliography}{29}
\bibitem{logic} In this respect there is an essential difference between the Periodic Anderson Model with a nearest-neighbor and a
local hybridization. In the former case (this work) the main drawback of the Hubbard-III method (i.e., the violation of Fermi-liquid
properties) becomes irrelevant for $U\gtrsim U_{c}$, due to the absence of significant low-energy spectral weight in the $f$-DOS. As 
a consequence, the qualitative predictions of the Hubbard-III method are expected to be reliable for all $U>U_{c}$. In particular, 
the prediction of a semimetal state is expected to be robust, since for $U>U_{c}$ there is no low-energy scale left where the
semimetallic state could break down. On the other hand, for a local hybridization (Ref.\ \onlinecite{ConGus}) there is significant 
low-energy spectral weight at large interaction and, accordingly, the Hubbard-III method is less appropriate. In this case, it is 
well-known\cite{ConGus} that the Hubbard-III approximation fails to predict the Kondo resonance and a hybridization gap.
\end{thebibliography}
\end{document}